**Ante, L., Saggu, A., Schellinger, B., Wazinksi., F.P. 2024. Voting Participation and Engagement in Blockchain-Based Fan Tokens. Electronic Markets 34 (26).**

# Voting Participation and Engagement in Blockchain-Based Fan Tokens

This Version: December 3, 2023

Lennart Ante [a]*

*[a] Blockchain Research Lab*

Aman Saggu [ab]*

*[a] Blockchain Research Lab*

*[b] Business Administration Division, Mahidol University International College*

Benjamin Schellinger [a]

*[a] Blockchain Research Lab*

Friedrich-Philipp Wazinski [c]

*[c] Católica Lisbon School of Business & Economics*

**Abstract:** This paper investigates the potential of blockchain-based fan tokens, a class of crypto asset that grants holders access to voting on club decisions and other perks, as a mechanism for stimulating democratized decision-making and fan engagement in the sports and esports sectors. By utilizing an extensive dataset of 3,576 fan token polls, we reveal that fan tokens engage an average of 4,003 participants per poll, representing around 50% of token holders, underscoring their relative effectiveness in boosting fan engagement. The analyses identify significant determinants of fan token poll participation, including levels of voter (dis-)agreement, poll type, sports sectors, demographics, and club-level factors. This study provides valuable stakeholder insights into the current state of adoption and voting trends for fan token polls. It also suggests strategies for increasing fan engagement, thereby optimizing the utility of fan tokens in sports. Moreover, we highlight the broader applicability of fan token principles to any community, brand, or organization focused on customer engagement, suggesting a wider potential for this digital innovation.

*Corresponding author: Lennart Ante, Blockchain Research Lab, Blockchain Research Lab gGmbH, Weidestraße 120b, 22083 Hamburg, Germany, Email address: ante@blockchainresearchlab.org

**Corresponding author: Aman Saggu, Business Administration Division, Mahidol University International College, 999 Phutthamonthon Sai 4 Rd, Salaya, Phutthamonthon District, Nakhon Pathom 73170, Thailand, Telephone: 0066 27005000, Fax: 0066 24415091, Email address: asaggu26@gmail.com; aman.sag@mahidol.edu.

# Introduction

Technological innovation, shifting fan demographics, and the evolution of global sports sectors have remarkably transformed the landscape of fan engagement (Frevel et al., 2022; Otto et al., 2011). Historically, fan engagement was relatively straightforward: maximization of game attendance, sports broadcast viewership, and team merchandise sales (Funk and James, 2001). However, the ascent of digital platforms and the ubiquity of social media have ushered in a new era of interactivity and community formation. Fans no longer merely spectate; they actively participate, share opinions, and potentially sway club decisions (Filo et al., 2015; Hambrick et al., 2010; Popp and Woratschek, 2016).

Fan tokens are an innovation forged by the convergence of blockchain technology, financial markets, and fan engagement (Ante et al., 2023; Zarifis and Cheng, 2022). This novel integration enables opportunities for heightened fan interaction and influence, fundamentally reshaping the realms of sports management, marketing, and the essence of fan engagement itself. In this study, we explore fan tokens, uniquely focusing on how fans interact with polls that allow them to vote on specific issues. The study begins by explaining factors of interaction, subsequently analyzes how voters behave, and then analyzes differences in fan engagement across different sports. Moreover, this study situates these findings within the broader digital innovation framework in the sports industry, offering a comprehensive and engaging perspective on this rapidly evolving landscape.

Fan tokens, digital crypto assets designed to amplify the involvement of supporters in professional sports teams, have garnered considerable attention since 2020. Numerous sports clubs have launched fan tokens, arguably making significant strides in the fan engagement space. For instance, the FC Barcelona token launch met with tremendous funding success (Cohen, 2020), while the Italian national team also experienced a rapid sell-out of its fan tokens (Marca, 2022). The industry-leading fan token platform, Socios, has contributed significantly to this trend, introducing official fan tokens for numerous club competitions and partnering with key industry figures, such as Lionel Messi or the UEFA club competitions (Socios, 2022a, 2022b). From a conceptual point of view, fan tokens offer their holders the ability to partake in team-related activities, such as voting on entrance songs to be played at the stadium or voting on the best player in a match. Fan tokens further have a speculative function due to being tradable as crypto assets on secondary markets and leveraging blockchain technology to ensure efficiency and trust (Ante et al., 2023). Theoretically, these assets offer fans increased engagement, interaction, and governance with their favorite sports clubs. Research into fan token consumption has uncovered the complex role of fan tokens for consumers, including identity, engagement, and financial motivations (Manoli et al., 2024).

Although fan tokens and the stocks of sports ventures are both tradeable assets, their value propositions and the nature of voting rights differ significantly. Fan tokens do not represent actual ownership of a venture. Instead, their value is more closely related to the engagement opportunities they provide and the demand among fans for those exclusive experiences and voting rights on minor club matters (Ante et al., 2023). Although both stocks and tokens offer voting rights, the scope varies greatly. While shareholders in football clubs may have a say in significant corporate decisions during shareholder meetings, fan token holders are limited to making choices on relatively minor yet engaging aspects of club life. This nuanced distinction underlines the divergent objectives of football stocks and fan tokens: one is primarily a financial investment with potential profits, while the other enhances fan interaction and engagement without offering financial returns from the club profits. Research has uncovered that the two asset classes can be considered independent of the other (Ersan et al., 2022).

To date, no studies analyze fan token voting polls. Early research into fan tokens focuses on financial or speculative aspects (Ante et al., 2024; Demir et al., 2022; Ersan et al., 2022; Scharnowski et al., 2022; Vidal-Tomás, 2023), a conceptualization and definition of fan tokens (Ante et al., 2023), motives of fan token consumption (Manoli et al., 2024), or gambling-like aspects of fan tokens (Lopez-Gonzalez and Griffiths, 2023). Therefore, the academic field still lacks an analysis that addresses the intrinsic value and utility of fan tokens for fans.

The ever-evolving landscape of sports fandom and fan engagement has been a focal point of numerous academic studies (Huettermann et al., 2019; Osborne and Coombs, 2013; Yoshida et al., 2014). This corpus of research has primarily focused on transactional fan behaviors, such as game attendance and merchandise purchases, with more limited exploration of non-transactional behaviors indicative of deeper fan engagement and commitment (Funk and James, 2001; Underwood et al., 2001). Moreover, the existing body of research demonstrates a gap in examining fan engagement in the context of fan tokens. One exception is the single case study of Ante et al. (2023) that presents some high-level statistics on fan token polls and fan token users but focuses on the conceptualization of fan tokens – and not an analysis of actual fan token utilization (i.e., exercising voting rights or club governance). This study aims to bridge this gap in the literature by providing an in-depth examination of engagement in fan token polls, the impact of the underlying content of polls on fan engagement (e.g., the type of questions, number of answer options), and the variances in fan engagement across sports sectors.

The primary objective of this study is to analyze the level of fan engagement in fan token polls[1] and the resulting (psychological) underpinnings of fan loyalty to broaden our understanding of the new dimensions of fan engagement that fan tokens potentially introduce. In tandem, we investigate the relationship between different poll themes and fan engagement. Exploring how various types of content, or questions, facilitate or hinder fan engagement in token polls contributes to the literature on the role of content provision and information dissemination in shaping fan engagement in the digital era (Byon et al., 2013; Huang and Chen, 2018). Finally, this study aims to discern the variations in fan engagement across different sports sectors. By achieving these objectives, we advance upon the limited literature on fan engagement, fan loyalty, and the impact of digital innovation in sports.

Specifically, we pose the following research questions (RQs), which will be motivated in detail in the following section.

- **RQ1:** What is the degree of voter engagement and participation in fan token polls?

- **RQ2:** How do different poll themes facilitate or inhibit fan engagement in token polls?

- **RQ3:** How does the level of (dis-)agreement among voters in fan token polls impact voter engagement and participation?

- **RQ4:** How does fan engagement in token polls fluctuate across different sports sectors?

Drawing upon a dataset of 3,576 fan token polls relating to tokens issued by market-leading fan token platform Socios, we developed a holistic understanding of fan token voting behavior through quantitative methods. Two complementary measures assess voting participation: the percentage of circulating tokens that voted and the number of fan token holders participating in polls. To reduce the complexity of the large corpus of textual poll data and to uncover underlying themes, natural language processing (NLP) techniques were employed, particularly the Latent Dirichlet Allocation (LDA) method for topic modeling (Blei et al., 2003). This technique builds on prerequisites such as text preprocessing, dictionary creation, and topic modeling, ultimately identifying distinctive themes characterized by a set of contributing words. We identify that clubs mostly provided fans with superficial decision-making options, such as voting on the best player or best moment, not embodying core aspects of club governance, which raises questions if fan tokens are just a short-term hype (and club financing tool via the sale of tokens) or a long-term product for fan engagement and community building. Descriptive and regression analyses are employed to address the research questions. Our study unveils that the aggregate average participation rate in polls exceeds 1% of the circulating token supply, signaling a potential under-exploitation of fan token capabilities. However, additional findings suggest that fan token polls captivate an average of 4,000+ participants, representing an average of 50%+ of total holders, demonstrating their effectiveness as fan engagement instruments. This revelation points towards a significant engagement depth, where fan tokens act as a voting mechanism and a vital tool for fostering a more intense sense of community and belonging among fans. The high percentage of token holder participation suggests that, while the total quantity of tokens used in voting may be small, the tokens effectively mobilized engagement with a substantial portion of the fan base. We discern several meaningful fan participation and engagement determinants, including individual themes discovered through topic modeling, the degree of consensus or disagreement among voters, and variables at the poll, sector, country, and club levels. These determinants highlight the multi-faceted nature of fan engagement, suggesting that fan tokens have the potential to significantly influence the dynamics of fan interaction with sports teams and clubs.

---

[1] Figure A.1 in the appendix displays a screenshot of an exemplary fan token poll from Socios.com. This poll enabled holders of Apollon Limassol fan tokens to select the 2021/22 season kits based on three predefined answer options. Participants were limited to voting with up to 200 tokens to prevent centralization, regardless of the number of tokens in their possession. In total, 15% of all tokens voted, whereas a successful vote required at least 5% of tokens. The poll opened on Jan 07, 2021, and closed on Jan 10, 2021.

This study contributes to both theoretical and practical understanding of fan engagement in the sports industry through the lens of fan tokens. Theoretically, it challenges the conventional view of fans as passive consumers (Funk and James, 2006), proposing a paradigm shift towards viewing them as active contributors within a participatory fan culture, achieved by analyzing the role of fan tokens in fostering community formation, enhancing fan loyalty, and facilitating value co-creation in brand communities (Huettermann et al., 2019; Popp and Woratschek, 2016; Pritchard and Negro, 2001). The study also delves into group dynamics in blockchain-based voting, offering insights into how agreement and disagreement within fan token polls influence fan engagement. The analysis revealed that polls with high levels of agreement among voters tend to result in greater participation, suggesting that consensus fosters a stronger sense of community and engagement. Conversely, high disagreement within polls can also enhance engagement, indicating a nuanced impact of voter (dis-)agreement on fan participation.

Practically, the research provides actionable insights for stakeholders such as sports franchises and marketers, emphasizing the transformative potential of fan tokens in reshaping fan engagement dynamics. Our findings guide the strategic integration of fan tokens into fan engagement and relationship management strategies, underscoring their potential to foster stronger fan-team relationships. Furthermore, the study highlights the importance of aligning fan token initiatives with team branding and marketing efforts. It further suggests leveraging the participatory and democratic nature of fan tokens for enhanced fan experiences. Additionally, it points to the significance of crafting polls and fan engagement activities that strategically navigate the balance of (dis-)agreement to maximize participation and foster a vibrant community dialogue. Finally, it underscores the need for sports industry stakeholders to embrace technological advancements like blockchain to stay competitive in the evolving digital landscape.

The remainder of the paper is structured as follows: the first section presents the conceptual background and motivates the research questions. The subsequent section details the data and methodology adopted for the study. The proceeding section presents the empirical results. The following section discusses these results in light of the existing literature, identifying the implications of our findings for theory and practice. Finally, the last section concludes.

# Conceptual background and research questions

The intersection of digital innovation and sports presents a novel arena to redefine fan engagement and transform sports management (Pandita and Vapiwala, 2023). One enabler of this digital revolution is blockchain technology, a decentralized and secure digital ledger system (Nakamoto, 2008) that has manifested in the form of various application areas in the sports industry (Glebova and Mihalova, 2023; Schellinger et al., 2022; Stegmann et al., 2023). Blockchain-based fan tokens or non-fungible tokens (NFTs) offer sports franchises innovative ways to cultivate loyalty and monetize fan interaction, marking a significant shift in the sports-fan dynamic (Carlsson-Wall and Newland, 2020; Martha et al., 2023; Schellinger et al., 2022). They equip fans with a sense of ownership and belonging by allowing them to participate in decision-making processes that heighten involvement (Ante et al., 2023). Simultaneously, fan tokens have the potential to foster a sense of community among fans, a key driver of fan engagement (Westerbeek and Shilbury, 2003). Sharing knowledge about the team and engaging in social communication through token polls can potentially facilitate deeper connections within the fan community (Pritchard and Funk, 2006). Recent research shows that fan tokens experience fewer bubble periods compared to cryptocurrencies, non-fungible tokens (NFTs) and decentralized finance (Assaf et al., 2024).

Academic literature and the cryptocurrency industry generally recognize fan tokens as fungible utility tokens developed on blockchain platforms, making each token of a specific type identical and interchangeable (Socios, 2024; Binance, 2024). These tokens empower holders with voting rights on certain organizational issues, access to exclusive rewards, and interactive experiences with the issuing organization, thereby boosting fan engagement and participation. Yet, the fan token domain is dynamically evolving, leading to an unconventional perspective where fan tokens overlap with non-fungible tokens (NFTs). In this innovative viewpoint, fan tokens act as organization-specific NFTs designed for promotional activities and community engagement (Zarifs and Cheng, 2022). They leverage NFTs' unique characteristics of ownership to foster community involvement and brand promotion, diverging from traditional fungible advantages (Martha et al., 2023). This approach paves new avenues for monetization, supports athletes' rights, and enhances fan interaction (Glebova and Mihaľová, 2023). This study explicitly concentrates on fungible fan tokens, referring to them as fan tokens throughout the manuscript.

This study enhances understanding of how fan tokens can and already do reshape the sports sector and fan engagement by investigating the presented research questions. This exploration offers pragmatic implications for sports franchises, providing strategic insights for fan token utilization. The findings will assist sports franchises in strategizing and optimizing fan experiences in the digital era, ultimately contributing to the continued growth and evolution of the sports industry.

## Fan engagement and loyalty

Fan engagement has emerged as a critical success factor for sports franchises, impacting their brand, sustainability, and growth. The construct of fan engagement is multi-faceted, with diverse dimensions extending beyond mere transactional exchanges and encapsulating a range of behaviors and attitudes (Huettermann et al., 2019; Yoshida et al., 2014). A pivotal aspect of fan engagement lies in fostering fan loyalty, defined by a deep-seated attachment and dedication to a team. Loyalty involves long-term commitment and consistent support extending beyond sporadic interaction or fleeting interest (Funk and James, 2006). It is important to note that fan loyalty benefits fans, providing them with a sense of community and identity and is simultaneously crucial for sports franchises. Loyal fans are more likely to offer continued support, contribute to the team financially, and advocate for the team within their networks, thereby contributing to brand strength, sustainability, and growth (Abosag et al., 2012; Pritchard and Negro, 2001).

Funk and James (2001) characterized fan engagement with a focus on transactional behaviors, such as attending games, watching games on television, purchasing team products, reading sports magazines and newspapers, and listening to games on the radio. These behaviors, crucial for understanding fan loyalty and affiliation, primarily cater to the self-interests of the fans and their consumption patterns. However, fan loyalty often translates into non-transactional behaviors, representing a deeper, more nuanced level of fan engagement. Non-transactional behaviors encompass prosocial actions such as positive word-of-mouth, collaborative event attendance, helping other fans, and engaging in dialogue and sharing knowledge about the team or game (de Ruyter and Wetzels, 2000; Westerbeek and Shilbury, 2003). These behaviors demonstrate elevated degrees of connection and commitment, indicating fan willingness to go beyond self-interest and contribute to the broader fan community and the sports franchise. This form of engagement, closely tied to fan loyalty, fosters a stronger sense of community, enhances team identity, and deepens the bond between the franchise and its fans (Westerbeek and Shilbury, 2003).

Fan engagement and loyalty are not merely transactional but are deeply rooted in psychological constructs. It is helpful to draw upon established theoretical frameworks to understand fan loyalty in the context of fan tokens. In line with the Uses and Gratifications Theory (U&G), which posits that consumers are active agents who use media to fulfill needs or desires (Katz et al., 1973), fans may engage with fan tokens for diverse reasons, including the need for personal identity (expressing loyalty and affiliation with the team), social interaction (connecting with other fans), information (learning about the team strategies and decisions), and entertainment. Further, the sense of empowerment and participation offered by token polls can satisfy the desire for personal agency, a gratification not typically fulfilled by traditional media (Sundar and Limperos, 2013).

Social Identity Theory (Tajfel and Turner, 1979; Turner and Oakes, 1986) provides another lens to understand fan loyalty, suggesting that individuals derive a portion of their self-concept from their membership in social groups. Sports fans, for example, often see their fandom as an aspect of their identity, using it to differentiate themselves from others (Osborne and Coombs, 2013; Reysen and Branscombe, 2010). When sports fans feel a strong sense of connection and identification with a team, they exhibit loyalty and participate in supportive behaviors. This group identification can drive fan loyalty, where fans remain committed to their team, deriving personal self-esteem from their team's success and standing by the team even in the face of failure or difficulty (Boyle and Magnusson, 2007).

The Investment Model (Rusbult, 1980) further expands the understanding of fan loyalty by looking at it as a form of relational commitment. The model suggests that commitment to a relationship, whether interpersonal or, in this case, between fans and their team, depends on three factors: (1) satisfaction with the relationship, (2) the quality of alternatives, and (3) the investment made into the relationship. Satisfaction might come from the joy of watching games, a sense of community among fans, or team success. Alternatively, it could come from other forms of related entertainment – such as fan token polls. Investments are critical to the model, including the time, money, and emotional energy dedicated to supporting the team. As fans invest more into the team (through purchasing merchandise, attending games, participating in fan communities, purchasing fan tokens, or participating in fan token polls), their loyalty to the team increases (Cayolla and Loureiro, 2014). In this light, fan tokens represent a unique form of investment. They provide fans with a tangible way to invest in and connect with their team, possibly strengthening their commitment and loyalty. The role of fan tokens, thus, is not just transactional or functional but can also be symbolic and emotional.

In this context, fan token polls are not just mechanisms for voting but tools for fostering deeper fan engagement and loyalty, enabling fans to voice their opinions and contribute to team (strategic) decisions. However, the nuances of these interactions and the extent of engagement and loyalty they facilitate remain largely unexplored. This study, therefore, seeks to delve deeper into the extent and nature of fan engagement in token polls. It aims to uncover how this novel approach to fan interaction influences the dynamics of engagement and loyalty, how it varies across different sports franchises, and what implications it holds for the future of fan engagement, loyalty, and sports management. Thus, we pose the following research question:

**RQ1:** What is the extent of voter engagement and participation in fan token polls?

## The role of content and narratives

The content deployed by sports franchises is a valuable resource in building relationships and interactions with fans, creating a community, and increasing loyalty and commitment (McDonald et al., 2022; Pedersen et al., 2007; Santos et al., 2019). It can act as an information carrier and an emotional trigger that further solidifies psychological connection and loyalty to their teams (Filo et al., 2015).

A critical pillar of club content strategy is shared knowledge, i.e., the information and insights about the team available to fans. According to multiple studies, shared knowledge strengthens fan engagement and catalyzes loyalty (Fisher and Wakefield, 1998; Westerbeek and Shilbury, 2003). This shared knowledge can encompass diverse forms, including team news, match insights, player stats, historical records, and more. A systematic provision and service of such information can help fans develop a more profound understanding of their team, facilitating more informed discussions and deepening their sense of belonging (Byon et al., 2013; Huang and Chen, 2018).

Using narratives in the content can be an effective tool to bolster fan engagement. Engaging and emotive storytelling can provide valuable information and resonate with fans on a deeper emotional level, fostering a stronger sense of connection with the team (Leyton Escobar et al., 2014; Meng et al., 2015). Narratives and topics based on team heritage, core values, victories, or players can potentially benefit high fan engagement (Kim and Hull, 2017; Yoshida et al., 2014). For instance, a club might create a poll where token holders vote on a story they wish to hear next – whether it represents a detailed narrative about a historic match, an in-depth profile of a legendary player, or a match behind-the-scenes look that imbues fans a sense of ownership over the content but also ensures narratives align with their interests.

In the realm of fan tokens, the themes of the polls may hold the potential to shape fan engagement levels. Thus, we aim to examine how different poll themes can either boost or impede fan engagement and raise the following research question:

**RQ2:** How do different poll themes facilitate or inhibit fan engagement in token polls?

We are especially interested, for example, in whether specific themes exist that resonate more profoundly with fans and trigger higher engagement. In addition, fans may perceive some themes as less meaningful, leading to decreased engagement. By uncovering these dynamics, this study will offer crucial insights for franchises striving to optimize their fan token strategies and amplify fan engagement.

## The role of (dis-)agreement and group dynamics

The role of agreement and group dynamics in influencing engagement and participation is a significant factor within various social contexts, and sports fandom is no exception. Rooted in theories of group polarization and social comparison, the tendency of group members to move towards an extreme position following group discussions can significantly influence individual participation levels in decision-making processes (Isenberg, 1986; Myers and Lamm, 1976).

Group Polarization Theory posits that individuals often shift toward a more extreme position after group discussions (Myers and Lamm, 1976). This phenomenon is quite observable in sports fandom, where fans exhibit strong group identities (Wann and Branscombe, 1990) – which has also been found in the consumption of fan tokens (Manoli et al., 2024). For instance, when fans perceive their opinion aligns with the majority consensus of the live results of an ongoing token poll, they may feel encouraged to participate. This sense of agreement validates their positions and fosters a sense of belonging to the group, which is critical to fan engagement and loyalty (Wann and Branscombe, 1993). Similarly, in line with Social Identity Theory (Tajfel and Turner, 1979), fans with a strong sense of belonging to a group might, therefore, participate more in activities that underline their group identity, including fan token polls of their preferred club where the broader voting outcome will (likely) align with the individual fan beliefs.

High levels of disagreement or conflict within a group can discourage fan participation (Souza et al., 2014). Social comparison theory indicates that fans might be less likely to engage when they perceive their opinion as unpopular or in the minority (Festinger, 1954). Perceived social isolation can lead to feelings of exclusion, reducing willingness to participate in group activities (Hitlan et al., 2006). Thus, fan token holders may be less likely to participate in polls where they assume or already foresee that they are in the minority and that their vote will ultimately not be among the poll winners.

Fan tokens, in particular, provide a platform where fans can directly voice their opinions, monitor (preliminary) results in real-time, and influence predefined decisions. Thus, the (dis-)agreement level in fan token polls might significantly impact fan engagement and participation. Understanding these dynamics becomes imperative for sports franchises striving to maximize fan engagement through fan tokens. Therefore, we aim to examine how the level of (dis-) agreement among voters in fan token polls influences fan engagement and participation, translating to our third research question:

- **RQ3:** How does the level of (dis-)agreement among voters in fan token polls impact voter engagement and participation?

By uncovering these patterns, we hope to provide valuable insights for sports franchises to optimize their fan token strategies.

## Fan engagement across sports sectors

Research into the intricacies of fan engagement across different sports sectors has become a crucial focal point in sports marketing. The cultural context of the sports, the intrinsic nature of the sport, and the adoption and effectiveness of digital platforms used for engagement influence the variations in engagement (Hunt et al., 1999; James et al., 2002; McCarthy et al., 2014). Each sport is defined by its unique culture, traditions, and social implications, which can lead to diverse fan behaviors and levels of engagement. For example, while international sports like football attract a large and varied fan base, eSports tend to attract a younger, more digitally-savvy audience that engages primarily through online platforms and live streaming services (Hamari and Sjöblom, 2017; Jenny et al., 2017).

Furthermore, the nature of the sport itself, including aspects like its rules, gameplay strategies, competition levels, and historical contexts, can significantly influence the degree and type of fan engagement. For instance, frequent breaks in play during an American football game might allow for more extensive engagement through digital platforms during the game itself compared to more continuous sports like rugby (Hunt et al., 1999). Further, eSports games such as League of Legends have complex strategies and fast-paced action, often resulting in high-intensity engagement from fans who follow along in real-time, debating strategies and outcomes in online forums and social media (Pizzo et al., 2018; Qian et al., 2020).

Digital innovation has brought about a shift in fan engagement, with different digital platforms leading to varying engagement levels across sports sectors, traditional and digital alike (McCarthy et al., 2014). Emerging innovations such as fan tokens offer unique engagement mechanisms and interactive features, potentially altering traditional engagement patterns and even introducing new ones, especially in digitally-native sectors such as esports. The globalization of sports has allowed fans worldwide to follow and engage with sports that might not be traditionally popular in their regions, leading to a more diverse fan base for many sports, including esports, which has seen rapid international growth (Dashti et al., 2022; Hamari and Sjöblom, 2017; Hill and Vincent, 2006).

**RQ4:** How does fan engagement in token polls fluctuate across different sports sectors?

# Data and empirical approach

## Data collection and processing

Our primary dataset of 3,576 fan token polls was acquired through the API of the online platform FanMarketCap and spans all fan token polls from their inception until July 2023. The website offers a comprehensive array of information related to various fan tokens. We were able to access and download a wide-ranging dataset of fan token polls, including detailed information such as token names, their corresponding clubs or teams, voting statistics, and poll information/questions.

In addition to the data retrieved from the FanMarketCap API, we added information regarding the type of entertainment or sport associated with each fan token, with football club being the most common type (appearing 2673 times). The other types, in descending order of frequency, are fan token ecosystem (271; the SSU token, a broader fan token initiative of the Socio platform[2]), esports (197), motorsports (175), rugby (90), football national team (67), tennis (60), and mixed martial arts (43). Further, we collected data on the country of the club associated with the fan token. The most common country is Turkey (appearing 878 times), followed by Italy (689), the United Kingdom (291) and Brazil (253). These data points provide a broader context of the fan token sector of influence, whether it be football clubs, national football teams, rugby, esports, or others.

---

[2] SSU Loyalty Tokens are the proprietary reward tokens of the Socios.com application, allocated to users upon advancing levels by accumulating points within the platform. Fans can earn the tokens through active engagement, including predicting match scores and voting in team polls. Initial allocation occurs upon user registration. SSU Loyalty Tokens serve a dual purpose; they enable users to partake in official decision-making polls and are redeemable for various rewards, including unique experiences, within the Fan Rewards section of the application.

## Estimation of fan token voting participation

We rely on two measures of participation in fan token polls: (1) the percentage of circulating tokens that voted (*total percentage voted*) and (2) the number of fan token holders that participated in polls (*number of users voted*).

The histogram in Figure 1 illustrates the distribution of the total percentage voted across all polls in the sample. The distribution is right-skewed, with most of the data points gathered towards the lower end of the scale and fewer data points towards the higher end. This skewness indicates that most polls have a relatively low total percentage voted, with fewer polls achieving a high proportion of votes. On average, around 0.66% of the total percentage votes in fan token polls, but this measure shows substantial variability across different polls with a standard deviation of 1.02%. The range extends from 0% to approximately 12.53%. For our multivariate models, we rely on a log-transformed measure of the total percentage voted to account for the skewness of the non-transformed data.

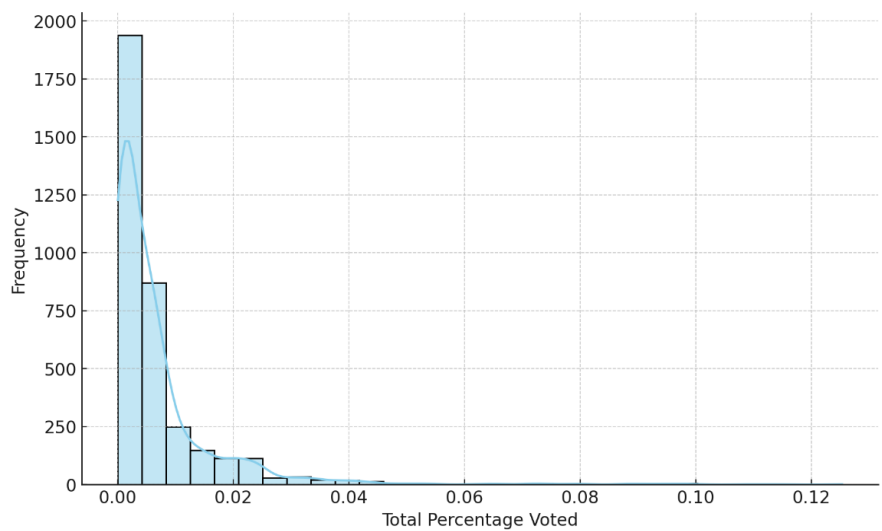

SSU Loyalty Tokens serve the same purpose as any other fan tokens; however, individual clubs, rather than Socios, issue polls. Therefore, these polls often concentrate on broader topics, such as leagues (e.g., "*Who was the Decisive Player of LaLiga Santander 2022-23?*") or events (e.g., "*Who will be the most missed player at the World Championship?*").

**Figure 1: Total percentage of tokens voted in fan token polls**

Fan token polls often restrict the number of tokens usable for voting. For example, if a user has 1 million tokens, only 50 can be used for voting. The token limit is associated with each poll, not the fan token itself. Thus, a fan token issuer can decide upon the token limit for polls they publish to their community. Therefore, while some polls have a predefined token limit of 1 (1 individual = 1 vote), others have a limit of, e.g., 50 or 100. Since our analysis includes polls conducted on the Socios platform, the source for this is the Socios platform or the poll data itself. While token limits prevent issues relating to centralization, a user could theoretically generate 20,000 wallets with 50 tokens each and achieve the same result. Thus, the total percentage voted measure should be significantly associated with the token limit. The average token limit across the sample is 53 (SD = 88) and extends from 1 to 1,000. The distribution skews towards lower values, with over half of the polls having a token limit of 1 (N = 2,420).

The boxplot in Figure 2 presents the distribution of the total percentage voted across different categories of token limit: 1, 2-50, 100 & 125, and 200. The 200 category includes a single observation with a token limit of 1,000. The categories represent different ranges of token limits. The total percentage voted is notably higher for the 100 & 125 token limit categories, with a median of 2.2%. Logically, the one token limit category shows a lower median of approximately 0.2%. The 2-50 and 200 categories show intermediate median values of around 1%. Notably, the spread of the total percentage voted (as indicated by the interquartile range) widens for the 100 & 125 token limit categories, suggesting a higher variability in voting percentages for this category. The results highlight the potential influence of token limit on voting behavior in fan token polls. Thus, the token limit is later included as a control variable in empirical analyses using the total percentage voted metric.[3]

---

[3] The FanMarketCap API classifies fan token poll types as either 'integrity' or 'engagement' but does not define these terms. Our analysis shows that polls labeled 'integrity' significantly correlate with the token limit (r = 0.84). This finding is logical because polls focused solely on engagement likely need fewer safeguards against voting fraud.

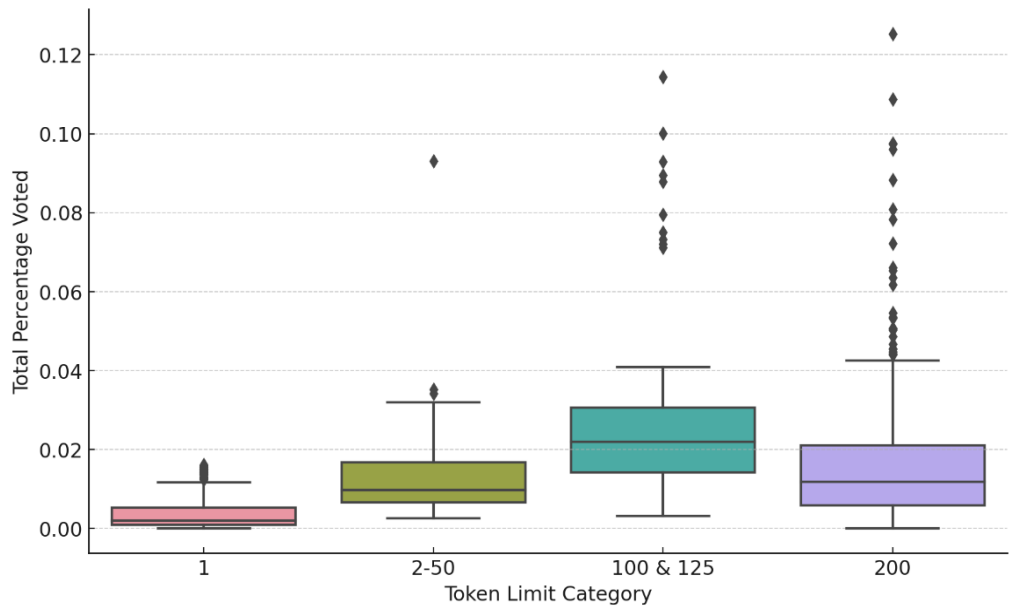

**Figure 2: Boxplot of total percentage voted by token limit category**

Based on the token limits, we derive how many people participated in a poll for all fan token polls with a limit of 1 – under the assumption that each fan token user only has one account. We focus on the token limit of 1, given the difficulties in identifying the degree to which individual users have fully utilized the token limit in any other poll.

Figure 3 displays the distribution of the number of tokens used for voting in polls with a token limit of 1. The mean of these polls amounts to 4,003 (SD = 2,025; median = 4,053), as shown by the peak of the distribution. However, there is substantial variability, as indicated by the wide spread of the distribution, with counts ranging up to approximately 23,000. Observations with high counts suggest that some polls experience much higher voting activity despite the token limit. This pattern underscores the potential influence of factors other than the token limit on voting activity in fan token polls.

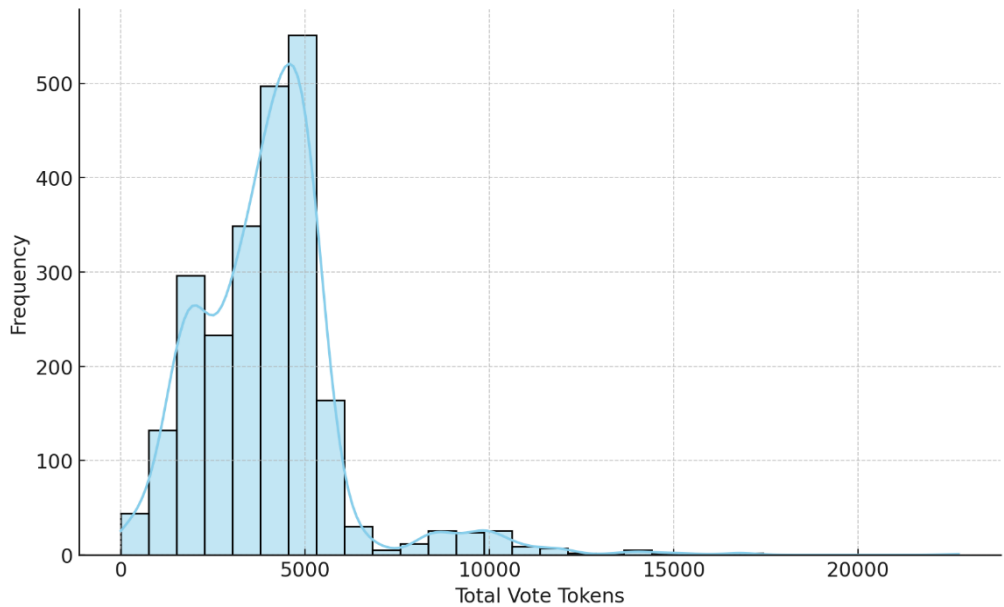

**Figure 3: Number of fan token holders voting in fan token polls.** The figure displays the distribution of the number of tokens used for voting in polls with a token limit of 1. In this setting, each vote equals one token holder.

A correlation analysis shows that the two metrics on participation are significantly negatively correlated ($r = -0.32$, $p > 0.05$). A noteworthy negative correlation exists between the total percentage of circulating tokens voted and the number of fan token holders participating in polls suggesting that a proportion of the token supply used for voting concentrates on fewer individuals. It further indicates a potential disparity in voting power, where a small group of holders may possess a significant portion of the tokens, thereby influencing decision outcomes more heavily than a larger, more diverse group of holders with fewer tokens each. This finding is meaningful as it raises questions about the true democratic nature of fan token voting systems and highlights potential issues in token distribution and voting equality.

## The role of (dis-)agreement in token polls

We collected the distribution of vote percentages, i.e., the percentage of votes allocated to a specific answer option, across all different answer options for each poll (available for $N = 2,720$). The average number of answer options per poll was approximately 7, with a median of 5 and a maximum of 57. Figure 4 presents the distribution of standard deviations of vote percentages across all polls, indicating (dis-)agreement or controversy among voters – and a potential driver of engagement. The majority of polls have a standard deviation between 0 and 20. It suggests that in most polls, the votes are concentrated around specific options, indicating a moderate agreement level among voters. The distribution peak is around 5-10, meaning that the most common scenario is for one or a few options to dominate the poll results while the other options receive fewer votes. However, the standard deviation is relatively high in many polls, up to around 60. It implies that in these polls, votes are evenly spread across various options, Indicating a high level of disagreement among voters. These polls could be considered controversial across participants. Table A.1 in the appendix offers an overview of the five polls with the highest and lowest standard deviation (highest and lowest disagreement).

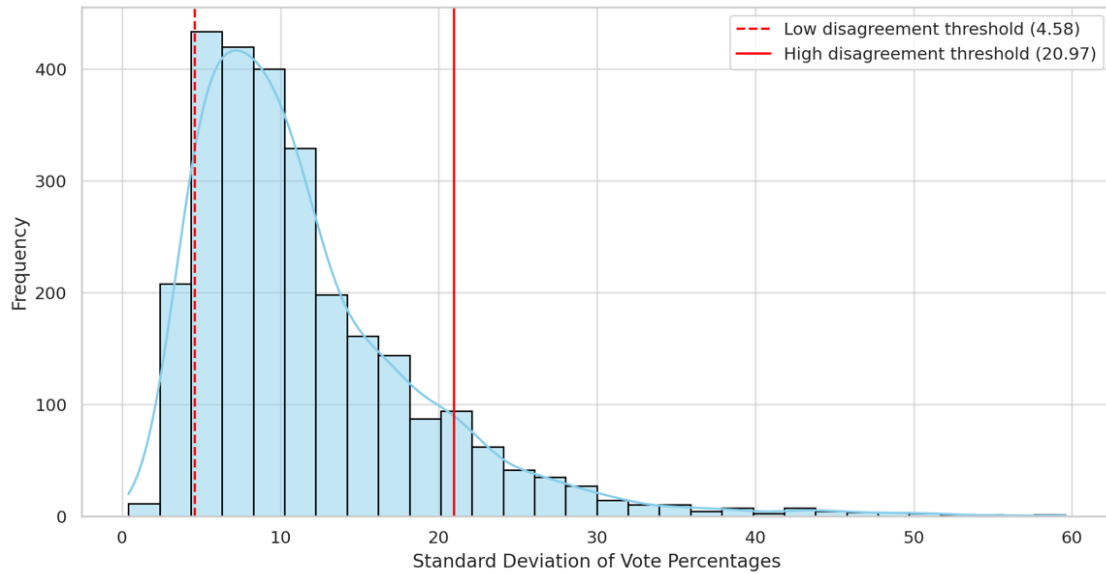

**Figure 4: Standard deviation of vote percentages across fan-token poll answer options.** The red lines indicate thresholds for the upper and lower 10% of the distribution.

In summary, while there is a tendency for vote percentages to be moderately concentrated around specific options in most polls, there is a wide variety of dispersion levels, highlighting the diversity in opinion among voters across different polls. The calculated standard deviation can quantitatively measure this diversity or controversy in our regression models in Section 4.2. The red dashed line represents the threshold for low disagreement (bottom 10% of standard deviations), while the red solid line represents the threshold for high disagreement (top 10% of standard deviations). For each of these two categories, we compute dummy variables.

Figure 5 visualizes the total percentage voted and total vote tokens (users) for the two dummy categories. In terms of the total percentage voted a t-test yielded a t-statistic of 1.97 and a p-value of 0.05, indicating a statistically significant difference between the two groups. Specifically, polls with low disagreement tend to have a higher percentage of votes cast than those with higher disagreement. For total vote tokens (limited to polls with a token limit of 1), the t-test showed a t-statistic of -4.85 (p >0.01), suggesting a comparatively significantly lower count of locked tokens in the low disagreement polls. These results indicate that the level of poll disagreement can impact both the total percentage of votes and the count of locked tokens and validate its inclusion in the multivariate models.

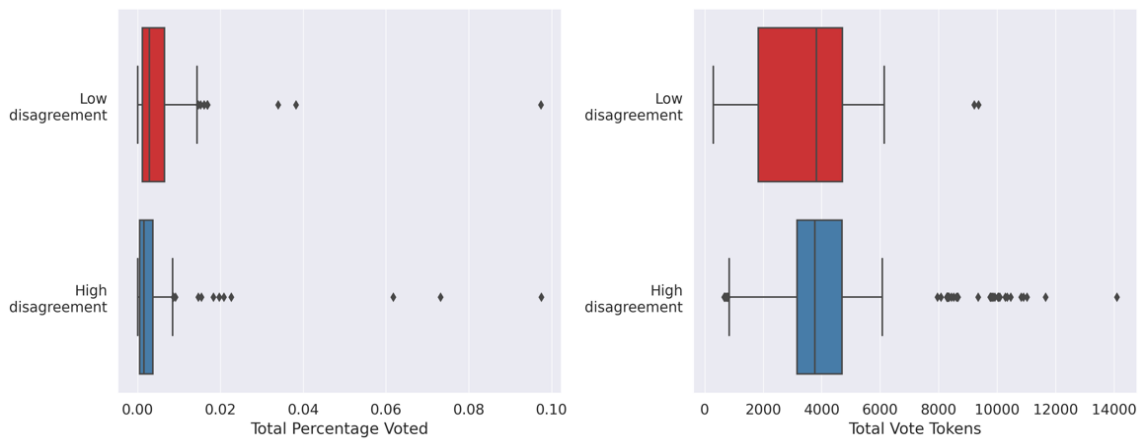

**Figure 5. Boxplot of total percentage voted and number of fan token voters by token polls with low and high levels of disagreement**

## Topic modelling

We adopted LDA for effective topic modeling since it provides an objective methodology to uncover underlying token poll themes using NLP. This approach allows for the automatic discovery of hidden topics within large volumes of text by assuming that documents are mixtures of topics and that topics are mixtures of words. The first step involved text preprocessing consisting of lowercasing all text to standardize the data and avoid word duplication due to case differences. In addition, we removed punctuation and stop words (e.g., the, that, on), which do not contribute significantly to topic meaning but can negatively influence model outcomes if not handled appropriately. Stop words are filtered out because they are abundant in the English language and provide little to no value in determining the context or themes of a document. We further refined the text by performing lemmatization, thus reducing words to their base or root form to simplify interpretation and enhance the model's ability to identify relevant topics (Blei et al., 2003).

The second phase entailed creating a dictionary of unique words in our dataset to map each word to a distinct identifier. Simultaneously, we generated a "bag of words" corpus, with each document represented by a list showing the frequency of occurrence of each word. A coherence score was computed for different numbers of topics to ascertain the ideal number of topics. The coherence score gauges the quality of each topic generated by the model, with higher values signifying superior quality (Blei et al., 2003). It calculates the similarity between the highest probability words within each topic, with the intuition that words that frequently occur in the same topic should have higher semantic similarity. Upon analyzing these scores, we selected the number corresponding to the highest coherence score – the score of ten.

Consequently, the LDA model resulted in ten distinct and uncorrelated topics that we classify into four main categories. Each topic corresponded to a set of words contributing most significantly to that topic's theme. The table below presents each topic with its top ten contributing keywords and their proportionate share in the dataset. By leveraging LDA's unsupervised machine learning capabilities, we successfully identified and labeled significant themes in the corpus, used for subsequent analysis. For each fan token poll, we derive a probability score indicating the degree of thematic overlap to each of the ten topics and generate dummy variables that take a value of one if a particular theme has the highest probability score of a fan token poll. This probability score is crucial for quantifying the relevance of each topic to a given document, facilitating a nuanced analysis of the content's thematic composition.

| Category | Topic | Top keywords | Share |
|---|---|---|---|
| Team and player highlights | 0 | season, race, rfk, 2021/22, jersey, weekend, fans, team, want, help | 4.75% |
| | 2 | match, player, goal, best, vote, mvp, win, guess, join, debate | 15.49% |
| | 3 | fan, player, month, token, choose, vote, holders, win, award, decide | 7.21% |
| | 5 | vs, best, mvp, season, vote, player, milan, game, goal, udinese | 11.97% |
| Game moments | 1 | moment, choose, match, la, song, pick, favourite, home, vote, warmup | 6.60% |
| | 8 | choose, win, laliga, best, matchday, favourite, chance, shirt, official, play | 9.45% |
| Design and fan interaction | 6 | choose, token, fan, holders, design, flag, corner, fc, pennant, match | 6.01% |
| | 7 | choose, design, one, want, like, see, fan, team, would, new | 17.03% |
| Visual and media engagement | 4 | photo, training, favorite, best, week, vote, last, match, fk, başakşehir | 16.75% |
| | 9 | team, favourite, f1, grand, league, prix, season, romeo, alfa, aston | 4.74% |

**Table 1: Identified themes via topic modeling.** The table shows categorized results of the Latent Dirichlet Allocation (LDA) method for topic modeling across 3,576 fan token polls. Top keywords of a particular topic and the relative share of a topic in the data.

# Results

## Descriptive statistics

Figure 6 presents two side-by-side boxplots comprising the total percentage voted and number of voters across various types of sports, ordered by the median total percentage voted from highest to lowest. Tennis and mixed martial arts have the highest median values in total percentage voted, whereas the fan token ecosystem has the lowest. It indicates a significant difference in voting participation across these types. The range of total percentage voted is broadest for mixed martial arts and motorsports, reaching up to around 12.5% and 11.4%, respectively, suggesting a high variability in the percentage of votes in these types. Outliers, represented by points beyond the whiskers, are present in several sports, notably mixed martial arts and motorsports, indicating occasional polls with unusually high or low voting percentages in these types.

The fan token ecosystem and motorsports have the highest median values by the number of voters, while rugby has the lowest. The fan token ecosystem's median is around 9,174, and rugby's is about 1,917, indicating substantial differences in voting participation across types – which are significant based on a t-test. The range of voter numbers is broadest for the fan token ecosystem, suggesting a high variability in the number of voters. These findings highlight the variability in fan voting behavior across different sports and fan token ecosystems. Some types elicit a consistently higher percentage of votes and have more voters than others, underscoring the importance of considering type-specific factors when analyzing fan voting patterns.

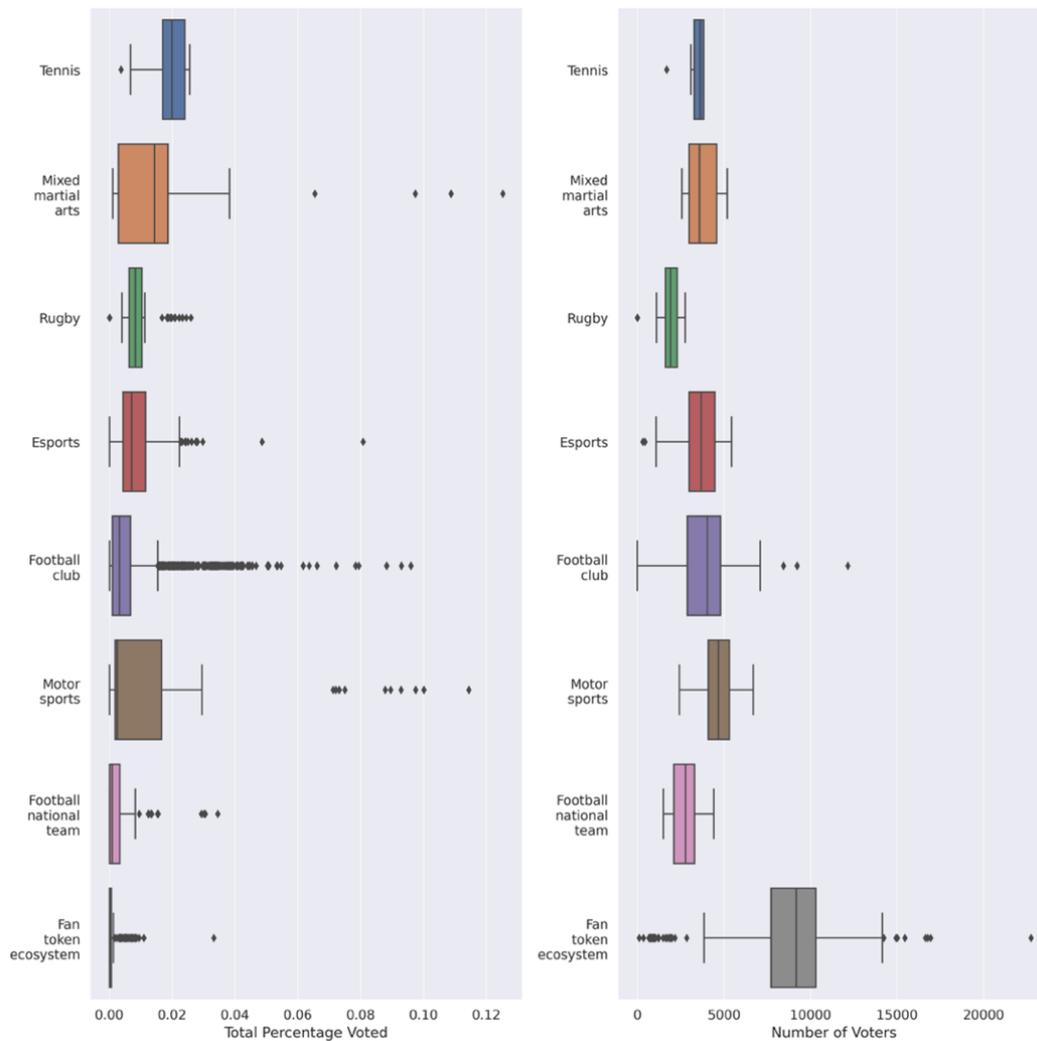

**Figure 6: Boxplot of total percentage voted and number of fan token voters by type of entertainment**

Figure 7 presents a comparative analysis of fan voting behavior across different countries of fan token ventures, leveraging the total percentage voted and the number of voters. Countries are systematically arranged in descending order based on the median total percentage voted, facilitating an intuitive comparison across the two boxplots. Poland has the highest median voting percentage (approximately 1.82%), while Portugal records the lowest median (close to 0.09%). Notably, Poland and Mexico exhibit wide interquartile ranges, implying substantial variability in voting percentages within these countries. Simultaneously, the boxplot illustrating the number of voters unveils distinct patterns. The international category tops the list with the highest median number of voters (8,619), indicating an expansive and engaged global fanbase associated with the Socios United Token. Mexico has the lowest median (1,960), indicative of comparatively lower fan participation. The international category also stands out for its significant variability in the number of voters, as demonstrated by its expansive interquartile range. The presence of outliers in multiple countries underscores the existence of sporadic polls with unusually high or low voting percentages or numbers of voters, potentially attributable to specific events or campaigns. These findings reveal the variability inherent in fan voting behavior across different countries. They suggest a complex interplay of factors, such as the popularity of specific tokens or the prevalence of fan token ecosystems influencing fan engagement across different geographical landscapes.

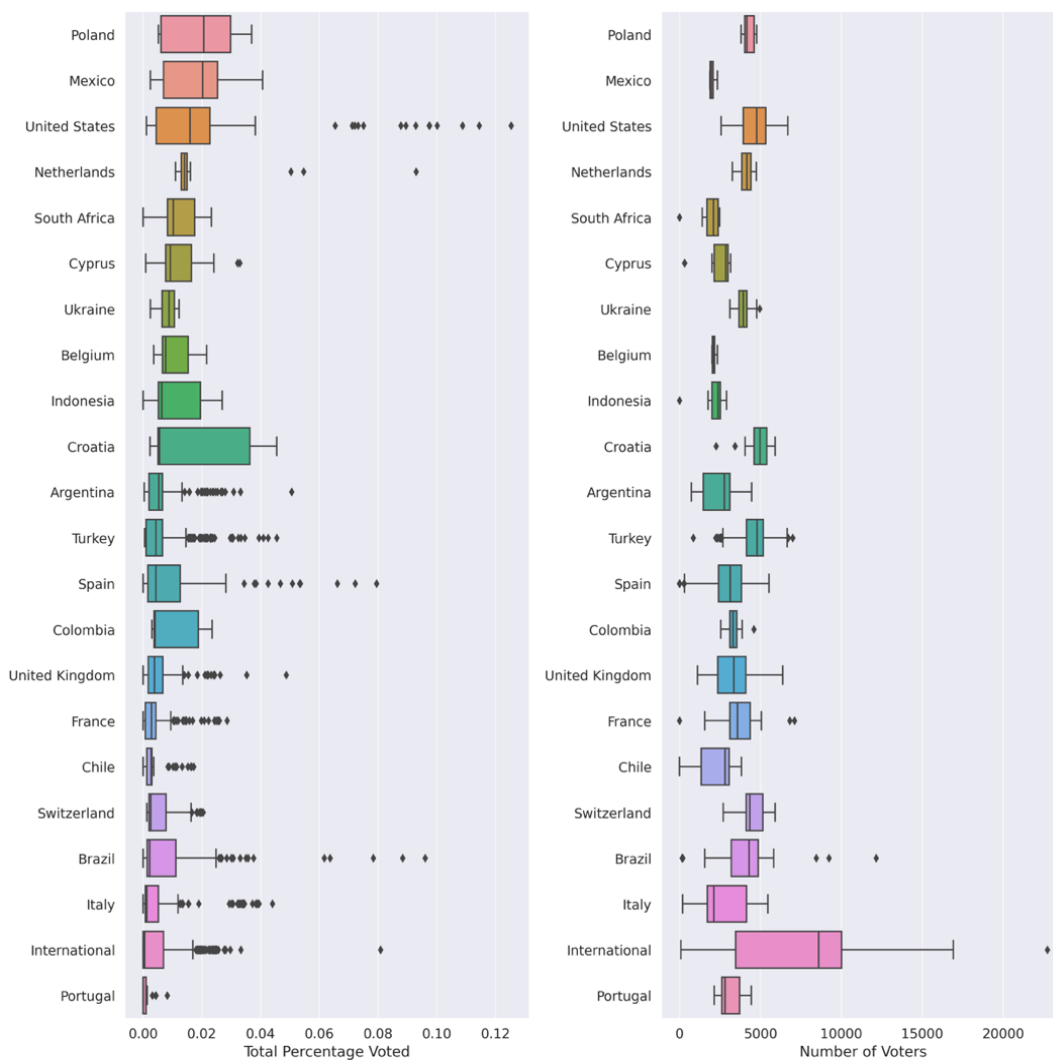

**Figure 7: Boxplot of total percentage voted and number of fan token voters by country of the fan token**

Figure 8 presents a boxplot of the total percentage voted and the number of voters across various fan tokens. The token with the highest median total percentage voted of approximately 3.93% is SE Palmeiras (sample size of 2), which also has a wide interquartile range extending from about 2.0% to 5.9%, indicating a wide spread of voting percentages. Interestingly, SE Palmeiras also has the lowest number of voters among the tokens – since the one poll with a token limit of one had few voters. On the other hand, Socios United has the highest median number of voters, with about 9,174 voters per poll, despite not being among the tokens with the highest voting percentages, suggesting a large and active community of voters but with a relatively lower percentage of total votes per poll.

Tokens such as Santos FC, Persija Jakarta, Portugal national team, S.S. Lazio, and Alpine F1® team have low total percentage voted, indicating lower voting activity. Tokens like Levante U.D. and Atlas FC show a wide interquartile range, indicating a higher degree of variability in the voting percentages for these tokens. Outlier identification for many tokens indicates occasional polls with unusually high or low voting percentages. These findings highlight the differences in fan voting behavior across different tokens, with some tokens consistently receiving a higher percentage of votes or having more voters than others. This variability underscores the importance of considering token-specific factors when analyzing fan voting patterns.

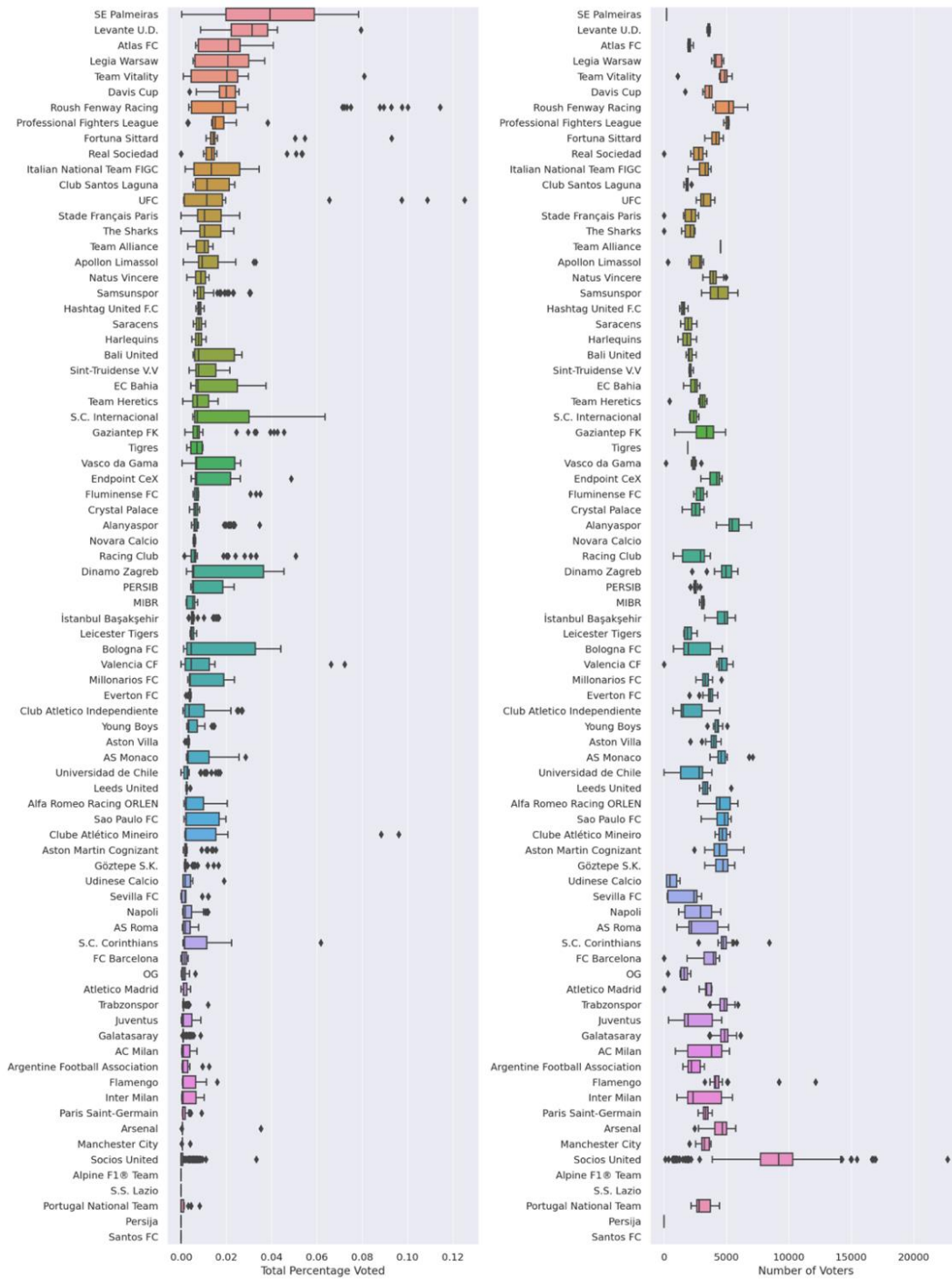

**Figure 8: Boxplot of total percentage voted and number of voters by fan token**

## Multivariate analysis

Tables 2 and 3 present results from multiple ordinary least squares regression models predicting fan token poll participation using different operationalizations of nine poll themes and unique poll characteristics and token themes. Formally, using vector notation, the logic of the following regression models can be described as:

$$Y = \beta_0 + \sum_i \beta_{i,Poll\ themes} X_{i,Poll\ themes} + \beta_{Token\ limit} X_{Token\ limit}$$

$$+ \beta_{Number\ of\ answers} X_{Number\ of\ answers}$$

$$+ \sum_j \beta_{j,Engagement} X_{j,Engagement} + \sum_k \beta_{k,Token\ themes} X_{k,Token\ themes} + \alpha_{Fan\ token}$$

$$+ \alpha_{Country} + \epsilon,$$

(1)

where, e.g., $\sum_i \beta_{i,Poll\ themes} X_{i,Poll\ themes}$ represents the summation of coefficients and variables for each poll theme, indicating that each theme I has its own coefficient and corresponding variable. Similarly, categories for token themes are includes as well as engagement metrics, which differ per model expression. $\beta_{Token\ limit} X_{i,Token\ limit}$ stands for the token limit. The $\alpha_{Fan\ token}$ and $\alpha_{Country}$ are the fixed effects controlling for unobserved heterogeneity within each fan token category and country. The models use probability scores for models (1) through (3) and dummy variables for best fit in models (4) through (6) and also for models in Table 3. Further, they rely on different engagement variables across models.

The adjusted $R^2$ values show that the models account for substantial proportions of variation in fan token poll participation, ranging from 63% to 75% across models from both tables. These high adjusted $R^2$ values, in conjunction with significant F-statistics, underscore the robustness and predictive power of our models.[4]

| | (1) | (2) | (3) | (4) | (5) | (6) |
| --- | --- | --- | --- | --- | --- | --- |
| | Coef. (SE) | Coef. (SE) | Coef. (SE) | Coef. (SE) | Coef. (SE) | Coef. (SE) |
| Poll themes | | | | | | |
| Theme 1 | -0.174 (0.153) | -0.396 (0.188)** | -0.371 (0.188)** | 0.018 (0.103) | -0.103 (0.120) | -0.088 (0.120) |
| Theme 2 | -0.139 (0.132) | -0.026 (0.142) | -0.022 (0.142) | -0.078 (0.094) | -0.022 (0.099) | -0.018 (0.099) |
| Theme 3 | 0.047 (0.147) | -0.050 (0.183) | -0.040 (0.183) | 0.057 (0.101) | 0.016 (0.124) | -0.007 (0.124) |
| Theme 4 | -0.168 (0.132) | -0.023 (0.142) | 0.050 (0.141) | -0.123 (0.095) | -0.047 (0.101) | 0.006 (0.101) |
| Theme 5 | -0.484 (0.143)*** | -0.318 (0.157)** | -0.389 (0.157)*** | -0.307 (0.099)*** | -0.128 (0.107) | -0.104 (0.107) |
| Theme 6 | -0.121 (0.154) | 0.155 (0.191) | 0.116 (0.192) | -0.141 (0.106) | -0.046 (0.129) | 0.067 (0.129) |
| Theme 7 | -0.307 (0.133)** | -0.441 (0.156)*** | -0.466 (0.156)*** | -0.160 (0.091) | -0.290 (0.105)*** | -0.301 (0.105)*** |
| Theme 8 | -0.036 (0.147) | 0.052 (0.162) | 0.081 (0.162) | -0.062 (0.098) | -0.056 (0.106) | -0.039 (0.106) |
| Theme 9 | 0.025 (0.165) | -0.007 (0.178) | -0.060 (0.178) | 0.030 (0.111) | 0.049 (0.061) | 0.035 (0.120) |
| Poll characteristics | | | | | | |
| Token limit | 0.008 | 0.007 | 0.007 | 0.008 | 0.007 | 0.007 |

---

[4] Correlation analysis and variance inflation factors (VIFs) reveal that all presented variables are suitable for regression analysis.

|  | | | | | | |
|---|---|---|---|---|---|---|
|  | (0.000)*** | (0.000)*** | (0.000)*** | (0.000)*** | (0.000)*** | (0.000)*** |
| No. of answers | -0.025 | -0.031 | -0.029 | -0.025 | -0.031 | -0.030 |
|  | (0.003)*** | (0.003)*** | (0.003)*** | (0.002)*** | (0.002)*** | (0.002)*** |
| Poll engagement | | | | | | |
| Level of agreement | | -0.017 | | | -0.017 | |
|  | | (0.003)*** | | | (0.003)*** | |
| Low Disagreement | | | 0.278 | | | 0.280 |
|  | | | (0.058)*** | | | (0.057)*** |
| High Disagreement | | | -0.143 | | | -0.154 |
|  | | | (0.063)* | | | (0.063)** |
| Token themes | | | | | | |
| eSports | 4.665 | 4.650 | 4.770 | 4.638 | 4.597 | 4.716 |
|  | (0.139)*** | (0.149)*** | (0.147)*** | (0.137)*** | (0.146)*** | (0.144)*** |
| Football club | 4.196 | 4.038 | 4.162 | 4.187 | 3.986 | 4.111 |
|  | (0.179)*** | (0.198)*** | (0.196)*** | (0.177)*** | (0.195)*** | (0.194)*** |
| Football nation | 3.968 | 3.245 | 3.414 | 3.986 | 3.255 | 3.423 |
|  | (0.278)*** | (0.282)*** | (0.282)*** | (0.241)*** | (0.283)*** | (0.283)*** |
| Mixed martial arts | 3.395 | 3.284 | 3.388 | 3.340 | 3.202 | 3.294 |
|  | (0.241)*** | (0.314)*** | (0.314)*** | (0.277)*** | (0.312)*** | (0.312)*** |
| Rugby | 5.351 | 5.308 | 5.500 | 5.349 | 5.225 | 5.422 |
|  | (0.205)*** | (0.228)*** | (0.225)*** | (0.203)*** | (0.225)*** | (0.223)*** |
| Motorsport | 3.839 | 3.851 | 3.926 | 3.791 | 3.764 | 3.840 |
|  | (0.219)*** | (0.236)*** | (0.236)*** | (0.214)*** | (0.230)*** | (0.230)*** |
| Tennis | 5.870 | 6.274 | 6.280 | 5.783 | 6.147 | 6.145 |
|  | (0.150)*** | (0.252)*** | (0.252)*** | (0.145)*** | (0.249)*** | (0.250)*** |
| Poll theme variables | Probabilities | Probabilities | Probabilities | Dummies | Dummies | Dummies |
| F-statistic | 0.000 | 0.000 | 0.000 | 0.000 | 0.000 | 0.000 |
| $R^2$(adj. $R^2$) | 0.74 | 0.76 | 0.76 | 0.75 | 0.76 | 0.76 |
|  | (0.74) | (0.75) | (0.75) | (0.74) | (0.74) | (0.75) |
| Number of observations | 3,355 | 2,569 | 2,569 | 3,355 | 2,569 | 2,569 |
| Fan token FE | Yes | Yes | Yes | Yes | Yes | Yes |
| Country FE | Yes | Yes | Yes | Yes | Yes | Yes |

**Table 2: Regression models predicting relative fan token poll participation.** The table shows the results of ordinary least squares regression models with heteroskedasticity-robust standard errors and constant (not reported) predicting fan token poll participation in percent. The variables (1) through (9) under 'poll themes' represent probability scores in models 1 through 3 and dummy variables for best fit in models 4 through 6. All models control for country- and fan token-fixed effects. Reference groups include the fan token ecosystem and poll theme 0. Statistical significance at the 1%, 5%, and 10% level are indicated by ***, ** and *.

|  | (1) | (2) | (3) | (4) |
|---|---|---|---|---|
|  | Coef. (SE) | Coef. (SE) | Coef. (SE) | Coef. (SE) |
| Poll themes | | | | |
| Theme 1 | 689.99 (328.20)** | 780.78 (328.61)** | 608.73 (205.11)*** | 663.17 (205.28)*** |
| Theme 2 | 93.48 (236.54) | 193.54 (237.13) | 288.46 (166.39)* | 350.61 (166.65)** |
| Theme 3 | 53.86 (349.85) | 121.15 (350.61) | 78.84 (229.74) | 116.39 (230.19) |
| Theme 4 | -82.57 (235.65) | 49.63 (235.71) | 74.54 (167.19) | 151.36 (176.45) |
| Theme 5 | -235.30 (274.87) | -146.63 (275.04) | 159.86 (186.53) | 227.59 (186.62) |
| Theme 6 | 42.53 (361.47) | 70.40 (362.30) | 151.84 (239.60) | 158.76 (239.98) |
| Theme 7 | -2,174.71 | -2,137.38 | -1,213.71 | -1,184.57 |
|  | (290.47)*** | (291.19)*** | (199.35)*** | (179.96)*** |
| Theme 8 | 1.028.23 (271.78)*** | 1,079.87 (271.66)*** | 869.97 (180.10)*** | 909.81 (179.96)*** |
| Theme 9 | 42.67 (305.12) | 100.74 (305.33) | 241.27 (209.07) | 265.21 (209.16) |
| Poll characteristics |  |  |  |  |
| No. of answers | -91.49 (6.03)*** | -82.02 (5.82)*** | -96.02 (5.90)*** | -86.21 (5.70)*** |
| Poll engagement |  |  |  |  |
| Level of agreement | -38.51 (5.34)*** |  | -39.52 (5.37)*** |  |
| Low Disagreement |  | 142.85 (97.17) |  | 126.57 (97.89) |
| High Disagreement |  | -750.77 (107.93)*** |  | -773.33 (108.58)*** |
| Token themes |  |  |  |  |
| eSports | -3,922.88 (880.75)*** | -3,843.12 (880.75)*** | -2,823.46 (418.86)*** | -2,633.94 (415.92)*** |
| Football club | -7,131.22 (1,332.29)*** | -6.820.87 (1,331.42)**** | -680.14 (727.24) | -489.82 (725.61) |
| Football nation | 5,881.34 (1,494.71)*** | -5,510.25 (1,494.00)*** | 478.10 (987.11) | 747.00 (985.27) |
| Rugby | -8,755.59 (1,229.78)*** | -6,470.91 (365.01)*** | -2,209.27 (837.58)*** | -1,864.76 (832.60)** |
| Motorsport | -5,291.70 (1,229.92)*** | -6,258.12 (1,401.76)*** | 993.91 (935.99) | 1,175.26 (934.69) |
| Tennis | -3.508.91 (400.29)*** | -3,516.52 (400.58)*** | -3,835.32 (396.07)*** | -3,837.71 (396.40)*** |
| Poll theme variables | Probabilities | Probabilities | Dummies | Dummies |
| F-statistic | 0.000 | 0.000 | 0.000 | 0.000 |
| $R^2$ (adj. $R^2$) | 0.64 (0.63) | 0.65 (0.63) | 0.64 (0.63) | 0.64 (0.63) |
| Number of observations | 2,297 | 2,297 | 2,297 | 2,297 |
| Fan token FE | Yes | Yes | Yes | Yes |
| Country FE | Yes | Yes | Yes | Yes |

**Table 3: Regression models predicting the number of fan token poll users.** The table shows the results of ordinary least squares regression models with heteroskedasticity-robust standard errors and constant (not reported) predicting fan token poll participation in terms of users. The variables (1) through (9) under poll themes represent probability scores in models 1 and 2 and dummy variables for best fit in models 3 and 4. All models rely on 2,297 observations and control for country- and token-fixed effects. Reference groups include the fan token ecosystem and poll theme 0 (fan favorite team). The mixed martial arts sector is excluded due to a lack of poll data with a token limit of 1. Statistical significance at the 1%, 5%, and 10% level are indicated by \*\*\*, \*\* and \*.

The results demonstrate that specific poll themes can significantly influence poll participation. Theme 5 consistently reveals a negative relationship with poll participation across all models, suggesting that this theme may discourage participation. The results for theme 1, potentially relating to votes on best moments, indicate a significant relationship across most models from both tables, a negative correlation in Table 2, and a positive and significant relationship with the number of poll users in Table 3. Initially counterintuitive, the results align with the negative significant correlation between the two measures of poll participation. However, theme 7 (e.g., new design) demonstrates negative associations with participation levels across most models. On the other hand, theme 8 (e.g., votes on best plays) indicates a significant and positive relationship with the number of fan token poll voters.

Among the poll characteristics, the token limit variable is significant across all models and positively associated with fan participation, indicating that a larger token limit may increase engagement – which is logical and underscores the relevance of the control variable. However, the number of answers shows a negative and significant association across all models, suggesting that greater numbers of choices may deter fan participation.

Poll engagement variables reveal that higher levels of either agreement or disagreement among fans may discourage participation, while polls with low disagreement levels may promote fan engagement. All token themes exhibit a statistically significant relationship with poll participation at the 1% level in all models from both tables, unsurprising, since the fan token ecosystem theme acts as reference group – which has the lowest vote percentage and highest number of voters (cf. Figure 7). eSports and football-related themes, in particular, substantially influence the number of poll users in the models presented in Table 3.

Table 4 reports the results from six distinct regression models intended to predict fan token poll (dis-)agreement, which includes two ordinary least squares regression models (models 1 and 4) and four logistic regression models (other models). The model logic, at a high level, can be described as:

$$
\begin{aligned}
Y = \beta_0 &+ \sum_i \beta_{i,Poll\ themes} X_{i,Poll\ themes} + \beta_{Token\ limit} X_{i,Token\ limit} \\
&+ \beta_{Number\ of\ answers} X_{Number\ of\ answers} + \sum_j \beta_{k,Token\ themes} X_{k,Token\ themes} \\
&+ \alpha_{Fan\ token} + \alpha_{Country} + \epsilon
\end{aligned}
\tag{2}
$$

which represents a variation of equation 1 that excludes the engagement variables – since they are now used as dependent variables.

In the poll themes, theme 2 (i.e., best player) demonstrates a significantly negative relationship with the level of agreement in models 1 and 4, indicating that polls themed around the theme are likely to see less agreement among fans. In contrast, the same theme shows a significant positive association with low disagreement in model 2, suggesting that this theme may lead to less discord among fan opinions. Thus, the actual underlying logic may not be linear but u-shaped. Themes 4 and 8 also reveal negative and statistically significant associations with the level of agreement in model 1, implying these themes may instigate more diverse fan opinions.

| | (1) | (2) | (3) | (4) | (5) | (6) |
|---|---|---|---|---|---|---|
| | Coef. (SE) | Odds ratio (SE) | Odds ratio (SE) | Coef. (SE) | Odds ratio (SE) | Odds ratio (SE) |
| Poll themes | | | | | | |
| Theme 1 | -1.377 (1.259) | 1.099 (1.654) | 2.054 (1.863) | -0.460 (0.794) | 0.735 (0.681) | 1.661 (0.944) |
| Theme 2 | -2.792 (0.935)*** | 7.178 (7.681)* | 0.942 (0.686) | -1.399 (0.656)** | 2.015 (1.239) | 1.259 (0.658) |
| Theme 3 | -1.325 (1.213) | 5.354 (6.233) | 3.265 (3.068) | -0.709 (0.822) | 1.899 (1.328) | 1.606 (1.012) |
| Theme 4 | -3.368 (0.924)*** | 1.322 (1.423) | 2.081 (1.592) | -1.660 (0.658)** | 0.543 (0.339) | 2.434 (1.366) |
| Theme 5 | -1.540 (1.073) | 1.550 (1.970) | 1.764 (1.408) | -0.846 (0.728) | 0.695 (0.542) | 2.123 (1.165) |
| Theme 6 | 0.574 (1.129) | 31.676 (39.687)*** | 4.502 (3.792)* | 0.213 (0.856) | 5.299 (3.906)** | 1.672 (0.986) |
| Theme 7 | -0.582 (1.033) | 20.515 (23.216)*** | 0.803 (0.566) | -0.431 (0.701) | 4.149 (2.678)** | 0.971 (0.478) |
| Theme 8 | -3.501 (1.062)*** | 2.096 (2.831) | 0.464 (0.329) | -1.948 (0.698)*** | 1.771 (1.225) | 0.762 (0.379) |
| Theme 9 | -1.318 (1.186) | 3.529 (4.255) | 2.501 (2.112) | -0.259 (0.797) | 0.602 (0.456) | 2.030 (1.135) |
| Poll characteristics | | | | | | |
| Token limit | -0.004 (0.002) | 0.999 (0.002) | 1.003 (0.001)** | -0.001 (0.002) | 0.998 (0.002) | 1.004 (0.014)*** |
| No. of answers | -0.456 (0.022)*** | 1.264 (0.288)*** | 0.372 (0.042)** | -0.449 (0.021)*** | 1.266 (0.029)*** | 0.379 (0.043)*** |
| Token themes | | | | | | |
| eSports | -8.750 (1.477)*** | 9.418 (14.196) | 0.038 (0.042)*** | -10.627 (1.364)*** | 12.844 (19.269)* | 0.052 (0.058)*** |
| Football club | -14.260 (5.621)** | 1.585 (3.281) | 0.0137 (0.023)*** | -13.539 (2.205)*** | 1.681 (3.466) | 0.024 (0.039)** |
| Football nation | -6.266 (-1.840)*** | 12.684 (35.552) | 0.002 (1.994) | -13.317 (3.595)*** | 28.159 (80.649) | 0.002 (0.906) |
| Mixed martial arts | -14.681 (6.121)** | 2.690 (8.428) | - | -15.778 (3.701)*** | 4.314 (13.609) | - |
| Rugby | -11.703 (1.523)*** | 12.346 (35.552) | - | -20.699 (2.768)*** | 10.749 (27.266) | - |
| Motorsport | -12.634 (5.836)** | 13.999 (39.704) | 0.002 (0.03)*** | -13.934 (3.297)*** | 12.782 (36.377) | 0.003 (0.006)*** |
| Tennis | -2.418 (1.609) | 8.666 (12.915) | 0.881 (0.674) | -2.091 (1.589) | 14.633 (21.741)* | 0.927 (0.705) |
| Dependent variable | Level of agreement | Low disagreement | High disagreement | Level of agreement | Low disagreement | High disagreement |
| Poll theme variables | Probabilities | Probabilities | Probabilities | Dummies | Dummies | Dummies |
| Regression model | OLS | Logistic | Logistic | OLS | Logistic | Logistic |
| F-statistic / Chi$^2$ prob. | 0.000 | 0.000 | 0.000 | 0.000 | 0.000 | 0.000 |
| R$^2$ (adj. R$^2$) / Pseudo R$^2$ | 0.40 (0.38) | 0.31 | 0.29 | 0.39 (0.37) | 0.32 | 0.28 |
| Number of observations | 2,569 | 2,306 | 2,160 | 2,569 | 2,306 | 2,160 |
| Fan token FE | Yes | Yes | Yes | Yes | Yes | Yes |
| Country FE | Yes | Yes | Yes | Yes | Yes | Yes |

**Table 4: Regression models predicting fan token poll (dis-)agreement.** The table shows the results of ordinary least squares (OLS) regression models with heteroskedasticity-robust standard errors. Logistic regression models include a constant (not reported) predicting fan token poll engagement in terms of (dis-)agreement in terms of vote distribution. The variables (1) through (9) under poll themes represent probability scores in models 1 through 3 and dummy variables for best fit in models 4 through 6. Reference groups include the fan token ecosystem and poll theme 0 (fan favorite team). Statistical significance at the 1%, 5%, and 10% level are indicated by ***, ** and *.

The token limit does display a statistically significant positive correlation with higher levels of disagreement while lacking significance in the other models. The number of answers variable reveals a significant negative relationship with the level of agreement in models 1 and 3 while positively related to high and low disagreement in the other models.

When considering token themes, there exist significantly negative relationships with the level of agreement for eSports, football clubs, football nations, mixed martial arts, rugby, and motorsport in model 1. In the logistic regression models 3 and 6, esports and motorsport manifest significant negative relationships with high disagreement, suggesting that these themes may lead to less discord among fans.

# Discussion

The central aim of this study was to delve into the characteristics of fan voting behavior within the fan token environment. Utilizing empirical analysis of polling data from a diverse range of geographies and encompassing a variety of fan token types and categories, we were able to shed light on the multifaceted nature of fan participation and engagement. This discussion serves to interpret our findings within the larger framework of fan engagement in token polls, integrating them with existing research, outlining their practical implications, and signposting potential areas for future research.

Our investigation into the first research question (RQ1)—*What is the degree of voter engagement and participation in fan token polls?*—has shed light on the multifaceted nature of fan voting behavior and how the operationalization of engagement affects the results. When looking at the 0.66% average measure of total voting percentages, the result could explain the failure of fan tokens to engage communities. However, the arguably low relative voting percentage is associated with its token limit. This limit ensures that a fan token holder can only use a certain number of tokens as input for a poll, thus decreasing the risk of fraud and centralization. For example, in July 2023, 6,323 token holders owned the 9.3 million circulating supply of the BAR token of FC Barcelona (Chiliz, 2023; RocketFan, 2023). The finding implies that, on average, each BAR token holder owned approximately 1,481 tokens. Thus, by definition, polls with a token limit of 1 could not exceed a total voting percentage of 0.06%.

However, when assessing users as an absolute measure, an average of 3,376 BAR fan token users participating in polls equates to a relative share of 53% of all token holders. The findings, therefore, can be interpreted as a success story of fan tokens. If one now argues that the FC Barcelona club, founded in the year 1899, has around 162,000 members (Barcelona, 2023), then a figure of over 6,000 (or 3,000 active) "digital members" since the launch of the BAR token in 2020 seems quite successful, even if they do not pay a membership fee in the traditional sense. In summary, judging the success of fan tokens or the level of engagement may be more appropriate using absolute fan token user numbers. The total percentage voted metric seems more suitable as a relative metric to generate comparability across different fan tokens and to assess longer-term, marked-wide developments.

Regarding our second research question (RQ2)—*How do different poll themes facilitate or inhibit fan engagement in token polls?*—our analysis yielded valuable insights. In assessing the poll themes—illustrated in Table 1, we found them to be relatively similar and somewhat superficial, not embodying core elements of club governance or decision-making processes. The most recurring themes revolved around subjective preferences such as new designs (17%), favorite photos from training sessions or matches (16%), or the best player or goal in a football match (15%). Arguably, these themes are primarily opinion-based and may not hold significant stakes in the actual governance of the sports club, i.e., the impact on broader club strategy or direction appears minimal. This further aligns with the findings that fan tokens represent engagement tools unrelated to governance or ownership as stocks of sports ventures represent (Ersan et al., 2022). Topics focusing on the player of the month (7%), the best moment or song for a match (6%), or the best play from a matchday (9%) are symbolic and lack significant strategic implications. While these themes equip fans with the ability to engage and express their preferences, their importance for substantive decision-making within sports organizations is arguably limited. These insights indicate a missed opportunity to engage fans in more substantial and impactful decision-making processes, enhancing their sense of ownership and commitment to the club.

Several key insights emerge from the regression analysis regarding fan engagement across different poll themes. Firstly, a theme potentially related to the best player showed variable engagement, hinting that, while integral to sports fandom, such topics do not consistently drive engagement in fan token polls. The theme centered on the best player or goal in the Italian league generally attracted less engagement, suggesting a limited geographical appeal or a smaller international audience interested in such narrowly defined discussions. Interestingly, polls centered around topics like new designs also saw less engagement, potentially indicating fan preferences for polls tied more directly to the sport or team performance. Finally, the best play theme demonstrated varied engagement; although generally not as engaging in terms of percentages, it attracted a higher absolute number of voters in certain instances. These insights highlight that fan engagement in token polls is not a straightforward process but a complex interplay of the poll theme, context, and presentation, and sports organizations should consider these dynamics to maximize fan engagement.

Our third research question (RQ3) – *How does the level of (dis-)agreement among voters in fan token polls impact voter engagement and participation?* – gave us a nuanced understanding of how agreement or controversy in fan token polls can significantly impact fan participation and engagement. Our findings revealed that polls with low disagreement, where votes concentrated around specific options, positively correlate with the total percentage voted and the number of users, whereas high disagreement exhibited a negative correlation. The role of agreement in enhancing voter participation corroborates the Social Identity Theory in sports fandom, where shared identities and consensus among fans can strengthen community ties and engagement (Manoli et al., 2024; Tajfel and Turner, 1979). The regression analysis reveals insights into the dynamics of fan engagement across different poll themes. For example, the "best player" theme demonstrated a statistically significant negative relationship with the level of agreement. While such topics are integral to sports fandom, they may also stir up diverse opinions and not consistently drive unanimous engagement in fan token polls. This variable also showed a positive relationship with low disagreement in the logistic regressions, indicating that, while it elicits diverse opinions, it may not be as polarizing.

Examining our fourth research question (RQ4) – *How does fan engagement in token polls fluctuate across different sports sectors?* – our findings have revealed some unexpected trends. Despite our previous indication that the fan token ecosystem had the highest median voter count (around 9,174), it does not reflect the highest voting percentages. Instead, sports such as tennis and mixed martial arts exhibit higher median voting percentages, showing greater relative fan engagement. The findings complicate the understanding of fan engagement, suggesting that a higher voter count does not necessarily equate to higher engagement. For instance, mixed martial arts and motorsports show a broad range of total percentages voted, up to 12.5% and 11.4%, respectively, signaling high variability in these sectors. Further, the regression models show that sports sectors consistently correlate, statistically significantly, with poll participation and engagement. This variability in fan engagement across sports sectors highlights the significance of sport-specific attributes in influencing fan engagement behaviors and preferences (Funk and James, 2006).

The fact that rugby, a globally popular sport, shows the lowest median voter count (about 1,917) suggests that the popularity or reach of a sector does not solely drive engagement levels, likely influenced by sport-specific factors and the conduct of fan token polls. These findings underline the importance of considering absolute numbers and percentages when assessing fan engagement across different sports sectors. Fan engagement is a complex interplay of factors that varies significantly across different sports and fan token ecosystems. Future studies on fan engagement should consider these dynamics to cultivate a more nuanced understanding of fan behavior.

## Implications for theory

In the conventional context, one typically sees fans as mere consumers of sports products and services (Funk and James, 2001). This study, however, suggests that fan tokens may shift this paradigm. Fan tokens create interactive opportunities that go beyond passive consumption of sports content, signaling the rise of a more participatory fan culture. This shift promotes the reimagining of fans as active contributors instead of passive consumers. Fan tokens can change how fans engage with their preferred teams, especially those distant and not part of fan groups. These tokens can transform fans from mere spectators to active participants, especially in activities that are not directly related to the game or match. This participation fosters community formation, enabling fans to connect, communicate, and bond over shared experiences. Consequently, the traditional consumer-oriented model may not adequately encompass these complex fan behaviors. We propose a fan token-expanded model of fandom that includes these community-oriented behaviors, offering a more comprehensive understanding of modern sports fandom.

Fan tokens are a critical element in brand community formation and maintenance. Popp and Woratschek (2016) highlighted that such communities develop around shared interests, a collective sense of identity, and intense member interaction. Fan tokens can significantly improve these interactions, strengthening community cohesion and loyalty. They create a symbolic connection between the brand (i.e., the sports club, team, or athlete) and community members, acting as an extension of the brand-community relationship that Muniz and O'Guinn (2001) emphasized. By doing so, fan tokens introduce a new dimension of engagement and loyalty, effectively expanding the relationship between customers and the brand.

The democratic potential of fan tokens, such as granting voting rights on club-related matters, i.e., the option of fan feedback sharing, can contribute to value co-creation in communities (Huettermann et al., 2019). This form of active engagement and ownership sense among community members is critical for community branding in sports (Pongsakornrungsilp and Schroeder, 2011). Therefore, fan tokens arguably serve as an innovative tool for facilitating value co-creation and fostering stronger relationships within brand communities. However, further empirical research is needed to fully understand the breadth of their impact on community-building and brand management.

While we cannot directly associate fan tokens with fan loyalty, our results suggest that fan tokens provide suitable instruments to engage fans, thus contributing to fan loyalty, underscoring the relevance of both social identity theory (Turner and Oakes, 1986) and the investment model (Rusbult, 1980). Fan tokens equip fans with a platform to express and reinforce their group identity, fostering a sense of belonging. Moreover, they symbolize an investment in the fan-team relationship, enhancing commitment and, thus, potentially brand loyalty.

Our investigation into the degree of (dis-)agreement within fan token polls and its subsequent influence on fan engagement contributes to the theoretical understanding of group dynamics in blockchain-based voting and governance. Our findings align with Wann and Branscombe (1993), illustrating that a sense of agreement often motivates individuals (fans) to participate, fostering their group identity. On the other hand, the greater engagement in terms of locked tokens in high disagreement polls underscores a complex side to fan behavior. This seemingly paradoxical behavior, which complements Jehn's (1995) work on group dynamics, necessitates a multi-dimensional understanding of fan engagement.

Finally, a crucial insight from our study relates to the transformative impact of blockchain technology, especially in the form of fan tokens, on the sports industry (Schellinger et al., 2022). This evolution calls for current theories to reconsider and integrate these digital trends, examining how technological advancements can revolutionize fan engagement and loyalty. As fan tokens become increasingly prominent, understanding their potential impact is paramount.

## Implications for practice

Our study's findings hold significant implications for sports franchises, marketers, and the broader sports and blockchain industries, offering actionable insights that can guide the practical application of fan tokens and other digital innovations in fan engagement strategies.

Firstly, our research highlights the transformative potential of fan tokens in reshaping fan engagement dynamics. Franchises can leverage these tokens to create interactive opportunities for fans, fostering a sense of empowerment and agency beyond mere consumption of sports content. By incorporating fan tokens into their digital engagement strategies, franchises can cultivate a participatory fan culture that encourages active involvement, community formation, and emotional connection (Mastromartino and Zhang, 2020).

Secondly, the enhanced gratifications associated with fan token participation highlight the importance of aligning these gratifications with the team's branding and marketing efforts. Fan tokens are not standalone entities but part of a comprehensive fan engagement strategy that delivers on the expectations and gratifications of the fans. Therefore, fan token launches should be carefully planned and effectively communicated, ensuring fan awareness of the benefits and opportunities associated with these tokens.

Moreover, the potentially significant role of fan tokens in strengthening fan engagement necessitates their strategic integration into fan relationship management practices. Fan tokens can be utilized as an investment platform that enhances commitment, fostering stronger fan-team relationships. This approach requires franchises to acknowledge and appreciate fans as key stakeholders and to design fan token strategies that consider the fans' interests and preferences.

While our study shows fan token polls with lower disagreement tend to attract more votes, it does not necessitate the avoidance of controversial topics that might incite higher disagreement. Although these may not increase voter participation, they could stimulate intense discussions, attracting more devoted fans and encouraging constructive debates, resonating with the broader literature on group dynamics where disagreement can spark innovation and community growth (Jehn, 1995). Therefore, sports franchises can strategically utilize these high-disagreement topics to enrich the fan token ecosystem and foster a more engaged, dynamic fan community.

Lastly, our study highlights the transformative impact of blockchain technology, particularly in the form of fan tokens, on the sports industry. It indicates the need for franchises and industry stakeholders to stay up-to-date with technological advancements and embrace these innovations in their strategies. Leveraging these digital trends can revolutionize fan engagement and loyalty, providing franchises with a competitive edge in the increasingly digital sports market.

## Limitations and future research

One primary limitation of this study is the reliance on quantitative data. While this approach provides a broad understanding of fan engagement patterns, it may not fully capture all the nuanced motivations and experiences of individual fans. Complementing this article's insights with qualitative methods such as in-depth interviews with fans or ethnographic studies in future research could significantly enrich the findings. Moreover, while we categorized poll questions broadly using NLP, future research could examine these in greater detail. Exploring different qualitative and quantitative models for poll question assessment may yield a more precise or distinct classification of poll themes, helping to uncover whether certain types of questions foster higher levels of engagement.

The correlation between fan token pricing and poll activity, the analysis of data on overall token holders, and unique wallets voting on each poll are promising avenues for further exploration. As fan tokens hold a monetary value that can fluctuate, it could be that these price dynamics significantly influence the motivation and capacity of fans to engage in polls. Understanding whether and how token price impacts fan engagement offers additional insight into the mechanics of this novel form of fan-sport interaction. Analyzing token holders and blockchain wallets could provide a more detailed understanding of fan engagement and offer insights into whether engagement is driven by a small, dedicated group of fans or more broadly distributed among the fan base. Ultimately, this approach will allow for a more nuanced and complete understanding of fan engagement in token polls.

The present study measures user counts for a share of polls where the token limit is set to one. Future research should extend this by incorporating data on overall token holders and unique wallets voting on each poll leveraging on-chain data. Such analysis would allow for a more detailed examination of fan engagement, offering insights into the extent of participation and the diversity of the participating fan base. For instance, analyzing unique wallets could show whether a small, dedicated group of fans or a broader distribution among the fan base drives engagement.

While comparing fan engagement across different sports sectors has yielded insights, future research could delve deeper into this area, investigating how fan engagement in token polls differs across teams within the same sport, possibly due to factors like team performance, player popularity, or regional factors.

# Conclusion

This article explores fan engagement through fan tokens, specifically examining fan token polls across various sports sectors, offering a novel perspective on digital innovation utilization within sports marketing, blockchain, and finance realms. The study demonstrates that fan engagement can be successfully leveraged through emerging technologies like fan tokens, as indicated by thousands of fan token holders regularly participating in polls. However, it is important to note that clubs currently grant fans only superficial decision-making powers, which may not sustain a long-term business model. Therefore, in order to grow long-term online fan token communities, clubs may have to delegate "more relevant" decision-making powers to fan token holders. This shift towards more significant decision-making could enhance the perceived value of fan tokens, deepening the engagement and loyalty of fans by making them feel like genuine stakeholders in their favorite clubs.

In the face of a digital transformation in sports fandom, fan tokens offer innovative engagement mechanisms, fostering unique interaction and participation patterns that extend beyond traditional fandom. This study establishes the potential of fan tokens to catalyze new fan bases in diverse sports sectors, allowing for the globalization and diversification of sports fandom. Despite the complexities and variations in fan engagement across sports sectors, our findings underscore the potential of fan tokens as a novel engagement tool. The insights of this study suggest that the strategic implementation of fan tokens could significantly impact fan culture, potentially leading to a more inclusive and engaged global fan community. However, as much as these insights are enlightening, they simultaneously highlight avenues for further investigation – from exploration of fan motivations to assessment of poll themes and relationships between token pricing and fan engagement.

The concept of fan tokens opens the door to exciting possibilities in various fields, transcending traditional engagement methods. This exploration into digital tokens reveals their potential to revolutionize interaction and loyalty in diverse sectors, from entertainment to retail. By applying the lessons learned from the sports sector, other industries could harness the power of digital tokens to foster community engagement, enhance brand loyalty, and create new revenue streams. While fan tokens have focused on sports, the underlying principles of these digital innovations promise broad applications, heralding a new era of digital engagement across multiple industries.

# Appendix

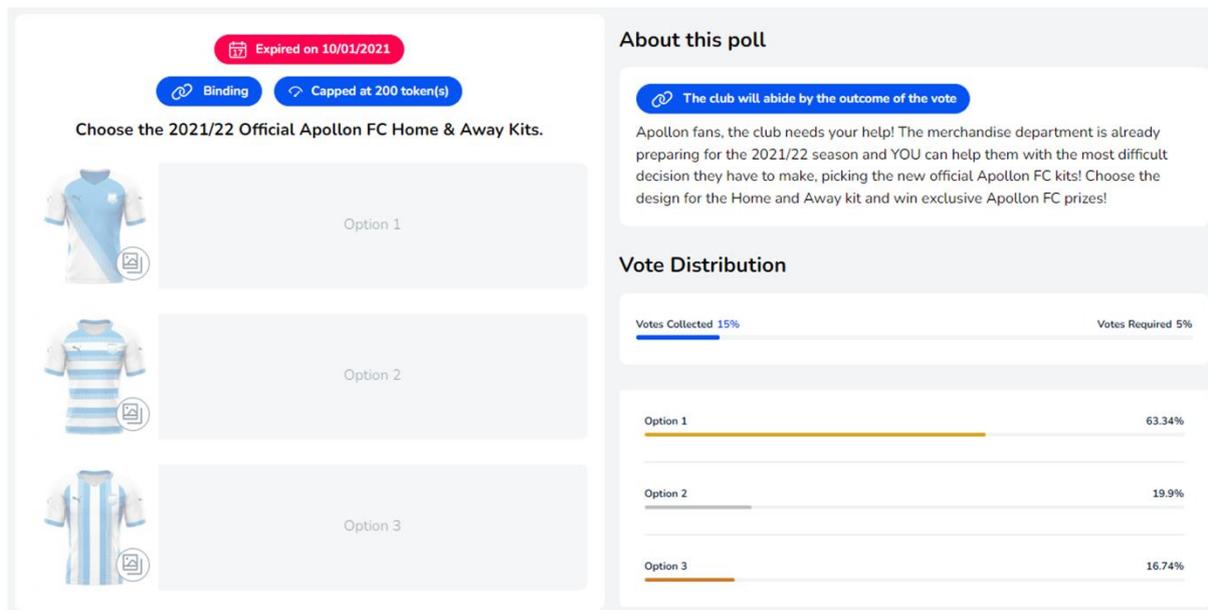

**Figure A.1: Example of a fan token poll on the Socios platform.** The screenshot from Socios.com shows the binding voting poll for Apollon FC's fan token holders to select the 2021/22 season kits.

| SD | Question | Answers (percentage) | Token | Date |
|---|---|---|---|---|
| Low disagreement | | | | |
| 59.57% | Pick the match review you want to watch on the Juventus official Facebook page on Sunday, May 10th at 15:00 CET. | 2015-16 Juventus-Sassuolo 1-0 (1.88%)<br>2017-18 Juventus-Milan 3-1 (92.12%) | JUV | May 20 |
| 55.09% | Should Benzema be called up by France for the EURO? | Yes (88.94%)<br>No (11.06%) | SSU | May 21 |
| 51.94% | Pick the match review you want to watch on the Juventus official Facebook page on Sunday, April 12th at 15:00 CET. | 2016-17 Juventus-Inter 1-0 (86.73%)<br>2017-18 Juventus-Roma 1-0 (13.27%) | JUV | Apr 20 |
| 51.64% | Are you happy to see Allegri back at Juventus? | Yes (86.52%)<br>No (13.48%) | SSU | May 21 |
| 50.75% | Choose the match review you want Juventus to stream on their official Facebook page on Wednesday, April 29th at 20:45 CET. | 2015-16 Juventus-Lazio 3-0 (85.89%)<br>2017-18 Juventus-Bologna 3-1 (14.11%) | JUV | Apr 20 |
| High disagreement | | | | |
| 0.40% | Among these 2 designs made by Jardel Lucas and Kenneth Anderson, which one do you prefer for the locker room at the training center? | Jardel Lucas (49.72%)<br>Kenneth Anderson (50.28%) | GALO | Nov 21 |
| 0.45% | Will Lionel Messi score FC Barcelona's first goal of the season? | Yes (49.69%)<br>No (50.31%) | SSU | Sep 20 |
| 1.13% | Which player will take over the Instagram account for 1 day? | Duje Čop (19.37%)<br>Luka Menalo (19.37%<br>Marko Bulat (21.89%)<br>Marko Tolić (20.15%)<br>Luka Ivanušec (19.22%) | DZG | Nov 21 |
| 1.24% | 3-2 win at El Cilindro against Aucas. Racing Fan Token holders, pick your favourite moment. [*pictures*] | ⚽ (34.19%)<br>5✖15 (31.91%)<br>🏆 (33.90%) | RACING | Apr 14 |
| 1.30% | Which competition will be a part of the next Team Heretics Barcelona Games? | 3x3 basketball (33.75%)<br>3x3 football (34.37%)<br>Chess match (31.88%) | TH | Jul 21 |

**Table A.1: Exemplary overview of fan token polls with lowest and highest levels of disagreement across answer options.**